\documentclass[manuscript]{aastex}

\slugcomment{Accepted for publication in the
  Astrophysical Journal.}

\shorttitle{Spitzer IRAC Imaging Survey}
\shortauthors{Carson et al.}

\begin{document}


\title{A Spitzer IRAC Imaging Survey for T Dwarf Companions Around M, L,
  and T Dwarfs: \\ Observations, Results, and Monte Carlo Population Analyses}


\author{J. C. Carson}
\affil{College of Charleston, Department of Physics \& Astronomy, 58 Coming St., Charleston, SC 29424, USA}

\author{M. Marengo}
\affil{Iowa State University, Department of Physics and Astronomy, A313E Zaffarano, Ames, IA 50011 USA}

\author{B. M. Patten}
\affil{Harvard-Smithsonian Center for Astrophysics, 60 Garden St., Cambridge,
  MA 02138, USA}

\author{K. L. Luhman\altaffilmark{1}}\affil{Pennsylvania State University, Department of Astronomy \& Astrophysics,
  525 Davey Lab, University Park, PA 16802, USA}

\author{S. M. Sonnett}
\affil{University of Hawaii, Institute for Astronomy, 2680 Woodlawn Dr.,
  Honolulu, Hawaii 96822, USA}

\author{J. L. Hora, M. T. Schuster\altaffilmark{2}}
\affil{Harvard-Smithsonian Center for Astrophysics, 60 Garden St., Cambridge,
  MA 02138, USA}

\author{P. R. Allen}
\affil{Franklin and Marshall College, Department of Physics and Astronomy, Lancaster, PA 17604, USA}


\author{G. G. Fazio}
\affil{Harvard-Smithsonian Center for Astrophysics, 60 Garden St., Cambridge,
  MA 02138, USA}

\author{J. R. Stauffer}
\affil{Spitzer Science Center, 1200 E California Blvd., Pasadena, CA 91106, USA}


\author{C. Schnupp}
\affil{Max-Planck-Institut f\"ur Astronomie, K\"onigstuhl 17, 69117
  Heidelberg, Germany}

\altaffiltext{1}{Center for Exoplanets and Habitable Worlds, The
Pennsylvania State University, University Park, PA 16802}
\altaffiltext{2}{Current Affiliation: Lincoln Laboratory, Massachusetts Institute of Technology, 244 Wood Street, Lexington, MA 02420}




\begin{abstract}
We report observational techniques, results, and Monte Carlo population
     analyses from a \textit{Spitzer} Infrared Array Camera imaging survey for substellar companions to
     117 nearby M,  L, and T dwarf systems (median distance of 10 pc, mass
     range of 0.6 to $\sim$0.05 $M_{\odot}$).  The two-epoch survey achieves
     typical detection sensitivities to substellar companions of [$4.5$
       $\mu$m] $\leq$ 17.2 mag for angular separations between about
     7$\arcsec$ and 165$\arcsec$.    Based on common proper motion analysis,
     we find no evidence for new substellar companions.  Using Monte Carlo orbital simulations (assuming random inclination, random eccentricity, and random longitude
     of pericenter), we conclude that the observational sensitivities translate
     to an ability to detect  600-1100K brown dwarf companions at semimajor axes $\gtrsim$35 AU, and to detect 500-600K companions at semimajor axes $\gtrsim$60 AU.      The simulations also estimate a
     600-1100K T dwarf companion fraction  of $<$ 3.4\% for 35-1200 AU separations, and $<$ 12.4\% for the 500-600K companions, for 60-1000 AU separations.  
\end{abstract}





\section{Introduction}
A variety of mechanisms have been proposed for brown dwarf formation.  For
example, brown dwarfs might form in a similar manner as stars, by the direct
collapse and fragmentation of molecular clouds.  While standard cloud
fragmentation theory has difficulty explaining the formation of such objects
(Boss 2001; Bate et al. 2003), modified versions, such as those including
supersonic magneto-turbulence, have the potential to produce the local high
densities required for brown dwarf formation (Padoan \& Nordlund 2004).  An
alternative theory is that photoevaporation by the ionizing radiation from
nearby massive stars prematurely halts the accretion of what would otherwise
be a more massive star, such as an F, G, K, or M type (Kroupa \& Bouvier 2003;
Whitworth \& Zinnecker 2004).  Finally, brown dwarfs could form through
instabilities in the outer ($\gtrsim$100 AU) disk around a star, and then be
pushed into the field by secular perturbations (Goodwin \& Whitworth 2007;
Stamatellos et al. 2007).  In this scenario, brown dwarf formation more
closely resembles planet formation than traditional star formation.  In
summary, the numerous models span a diversity of mechanisms, ranging from
those most similar to star formation to those analogous to planet formation.

Multiplicity may be a powerful tool to test some formation mechanisms, because
several theories argue that dynamical interactions may be essential for the
formation of brown dwarfs (Reipurth \& Clarke 2001; Boss 2001; Bate et
al. 2002; Delgado-Donate et al. 2003; Umbreit et al. 2005; Goodwin \&
Whitworth 2007; Stamatellos et al. 2007).  In Reipurth \& Clarke (2001) for
instance, the dynamical evolution of a group of protostars causes a member to
be ejected from the natal cloud core, before the member can otherwise reach
the Jeans mass necessary to form a star.  Such scenarios would make it difficult to produce wide-separation, ($>$20 AU) low-mass binaries.  However, several examples of wide binary
brown dwarfs have been discovered in recent years (Luhman 2004; Luhman
et al. 2009, references therein).  These observational discoveries, and others
like them,
therefore have the ability to constrain or perhaps disprove many ejection scenarios.        
  

While individual discoveries are intriguing, providing accurate multiplicity
statistics, in a consistent fashion, is necessary to rigorously test
predictions of formation theories.  A number of large-scale imaging surveys,
ranging from optical wavelengths through the mid-infrared, have been carried
out to help advance this effort (e.g. see Burgasser et al. 2007, Gelino et al.
2011, and references therein).  
 However, for the coldest temperature brown dwarfs ($<$800K), the 4.5 $\mu$m
 wavelength band of the Infrared Array Camera (IRAC; Fazio et al. 2004) on the \textit{Spitzer Space Telescope} (Werner et al. 2004) offers the best sensitivities of any currently available instrument, as long as the companion is well distinguished from the primary star's point spread function (PSF) (Patten et al. 2006).
Large-scale IRAC multiplicity surveys,  such as the multi-epoch study
described in this paper, provide a powerful tool for probing the multiplicity
of the coldest temperature objects.  In this regard, surveys like ours can
probe a phase space (cold temperatures, wide separations) inaccessible by
alternative surveys focused on near-infrared and optical imaging, Doppler
spectroscopy, or transit photometry.  Accompanying Monte Carlo population analyses allow for rigorous population statistics to be drawn from the observational results.  

Sections 2, 3, and 4 of this paper describe the target sample, observations, and data analysis, respectively, for this $Spitzer$ IRAC observational survey.  Section 5 summarizes the observational sensitivities.  Section 6 discusses the Monte Carlo simulations. Section 7 summarizes the investigation.

\section{Target List}
For our target list, we selected 117 nearby ($<50$ pc) bright M, L, and T dwarfs for imaging with IRAC. 
 The sample was designed to be a reasonably large collection of easily
 observable low mass (M, L, and T) stars and brown dwarfs that would allow one to constrain the frequency
 of wide, low-mass binaries.  Out of the 117 targets, 31 are known multiple
 systems (see Table 1 for individual references).  This value includes resolved and unresolved multiple systems as
 well as  single stars with known narrow-separation planets, as determined
 from Doppler spectroscopy surveys.   The observations were carried out between 2003 and 2009.
Table 1 summarizes the target spectral types, $Spitzer$ program identification
(PID) numbers, distances (from parallax and spectral type fitting), and
multiplicity information.  The primary stars consist of 52 M dwarfs, 37 L
dwarfs, and 28 T dwarfs.  The distances range from 1.3 pc to 43.8 pc with a
median distance of 9.9 pc.  A rigorous calculation of the mass range of these
primaries requires information on both target spectral type as well as age,
and our system ages are largely unknown.  However, a Bayesian analysis of
local solar neighborhood late-M, L, and T dwarfs by Allen et al. (2005)
reports mean masses as a function of brown dwarf spectral subtype, based on
Burrows et al. (2001) evolutionary models.  If we apply those statistical
trends to our target list, along with Baraffe \& Chabrier (1996)
mass/spectral-type relations for the early M dwarfs, we estimate that our
primaries have a mass range of 0.6 to $\sim$0.05 $M_{\odot}$.  As shown in
Table 1, about 30\% of the targets have no published distance measurements.
For these targets, we approximated a distance based on the 4.5 $\mu$m magnitude,
spectral type, and magnitude-spectral-type trends reported in Patten et
al. (2006).  The
magnitude-spectral-type  technique results in typical
distance errors of around 10\%, when tested on target stars with known
parallaxes.  The fitted distance might be unreliable for individual targets,
because unresolved binarity can affect the magnitude for a given spectral
type.  However, over a large ($>$100) target population, they are adequate for
delivering accurate statistics on the distance distribution.  Hence, distance uncertainties should have no significant adverse effects on our Monte Carlo companion statistics.  

     The target selection criteria used to plan the original PID 33 and 35 observing programs have the following biases: 1) a focus on relatively bright sources that accommodate high signal-to-noise (S/N) photometry of the primary; 2) an emphasis on relatively uncrowded fields; 3) a priority toward targets with known trigonometric parallaxes; and 4) an effort to avoid small-separation ($\leq$6$\arcsec$) binaries.  Section 6 discusses how these biases might affect our multiplicity statistics.
       
The first epoch observations of the 117 targets were performed in $Spitzer$
programs 33 and 35.  Second epoch follow-up observations were carried out in PID 50013,
30298,  60046,  3736, 70021, and 30179 programs.  The motivation for focusing
on multi-epoch targets was the capability of using common proper motion as a
measurement for bound companionship.  Only one of the targets from the PID 33
and PID 35 programs, 2MASS J17281150+3948593, lacked a second epoch
observation.  Three targets from our original list, DENIS
J153941.8-052042, 2MASS J07171626+5705430, and 2MASS J03480365+2344114,  had
second epoch observations, but with insufficient proper motion to enable an
astrometric test for bound companions.   We decided to discard 2MASS
J03480365+2344114, due to significant controversies in its distance and
multiplicity (Kirkpatrick et al. 1999), as well as the inadequate second epoch
measurements.    We retained the three remaining non-ideal targets in order to
keep the original target sample as intact as possible.  To
identify T dwarf companions using only one epoch of IRAC data, we used
photometry in four IRAC bands to perform a color-color identification (see
Luhman et al. 2007, Marengo \& Sanchez 2009).  Such an analysis is usually
limited by the IRAC 5.8 $\mu$m sensitivities, due to a combination of
high thermal background as well as a declining T-dwarf brightness at this wavelength   (Patten et al. 2006).  Common proper motion
analysis only requires detections in one photometric band, so we have the
advantage of focusing on the 4.5 $\mu$m region, which is the most sensitive IRAC band for T-dwarf detection.  This equates to an effective sensitivity floor $>$ 2
magnitudes deeper (at 4.5 $\mu$m), as determined for targets with and without
common proper motion data.  The superior 4.5$\mu$m sensitivities derives from a combination of low thermal background and a maximum brightness for late T dwarfs (see Patten et al. 2006).


\section{Observations}


The 117 targets were observed between 2003 and 2009  with the IRAC camera on
the \textit{Spitzer Space Telescope}.  The IRAC plate scale and field of view
are $1.\arcsec$2/pixel and 5.$\arcmin$2 $\times$ 5.$\arcmin$2, respectively
(Fazio et al. 2004).  The IRAC full width at half maximum (FWHM) at 4.5 $\mu$m
is 1.$\arcsec$6.  With the exception of 2MASS J17281150+3948593, all targets
were observed in at least two epochs.  Of the the two-epoch targets, all but
DENIS	J153941.8-052042 and 2MASS J07171626+5705430 exhibited $\geq$
0.$\arcsec$5 of proper motion between observations.  A motion of 0.$\arcsec$5, equivalent to about one-third of a FWHM, was the minimum shift in a point source's PSF where we found we could confidently check for co-movement with the primary, the test for a physically bound companionship.   
The first epoch observations were collected in all four IRAC channels (3.6, 4.5, 5.8, and 8.0 $\mu$m) while the second epoch observations were taken only at 4.5 and 8.0 $\mu$m.  For the common proper motion analysis, we focused only on the 4.5 $\mu$m data, because of its superior sensitivity for late T dwarf detections.  The observations at 4.5 $\mu$m for PID 33, 35, 50013, 30298, 70021, and 60046 were performed in the manner described by Patten et al. (2006).  For the cases of 2MASS J17281150+3948593, 2MASS J07171626+5705430, and DENIS	J153941.8-052042 (all from PID 35), where proper motion tests were not possible,  we used data collected in all four IRAC bandwidths, as described in Patten et al. (2006), to perform a color-color identification of possible substellar companions.  Data from PID 30179 was used for 2MASS J15344984-2952274, 2MASS J07271824+1710012, GJ 570 D, and 2MASS J04151954-0935066.  These target observations from PID 30179 all used 26.8 second exposure times, net integration times of 214.4 seconds, and a 5-position dither pattern distributed within a 13.$\arcsec$2 radius.  
PID 3736 data, which included SDSS J125453.90-012247.4 and 2MASS J15074769-1627386, used a 10-point and 16-point dither pattern, respectively, distributed within a 143$\arcsec$ dither radius.  Individual frame times were 26.8 and 10.4 seconds, respectively, with net integration times of 536 and 197.6 seconds.     
About a dozen targets exhibited saturation in the 4.5 $\mu$m images.  The
saturation radius reached a maximum of 4$\arcsec$.

\section{Data Analysis}

\subsection{Image Reduction}
Basic data reduction for all the observations was performed with the Spitzer
Science Center (SSC) IRAC Pipeline (version S14.0.0), which produced Basic
Calibrated Data (BCD) frames and data quality masks for each individual
exposure.  We used the post-BCD IRACproc package (Schuster et al. 2006) to
obtain, for each target, a single flux calibrated mosaic combining all the
individual exposures in each epoch for a given filter.  Using a bilinear
interpolation, IRACproc mapped the final mosaics onto a pixel grid with
0.$\arcsec$86/pixel resolution.  The effective field of view for object
identification was 5.$\arcmin$2 $\times$ 5.$\arcmin$2.  IRACproc makes use of
the SSC mosaic software MOPEX and provides enhanced outlier (cosmic ray)
rejection.  Figure 1 shows an example of a reduced 4.5 $\micron$ image.

\subsection{Companion Identification}
For each target, we used the task {\it starfind} within IRAF to
measure astrometry for all point sources in the 4.5~$\mu$m images from
the first- and second-epoch observations.
We matched the resulting source lists between the two epochs and
used the average differences in right ascension and declination to correct
for any offsets in the coordinate systems of the two epochs (which
were usually less than $\sim0.2\arcsec$).
To identify possible companions to a given target, we checked for sources
with motions that were consistent with the motion expected for the
primary based on its known proper motion and the time baseline between the
two epochs. We also verified that the published proper motion of the primary
agreed with the value that we measured using the IRAC data.
For the 114 targets where a common proper motion analysis was possible, we did not identify any new companions to the primaries.

For 2MASS J17281150+3948593, 2MASS J07171626+5705430, and DENIS
J153941.8-052042, where suitable second epoch data were not available, we
searched for substellar companions based on a comparison of candidate
magnitudes and colors using all four IRAC bands.  The photometric analysis was
carried out in a manner described by Patten et al. (2006).  This included a photometry
aperture radius of 4 native IRAC pixels and a sky annulus extending from 10
to 20 native IRAC pixels.  These parameters were applied
for all IRAC bands.  We tested for a substellar identification using a
photometric $k$-NN analysis described in Marengo \& Sanchez (2009).  The
$k$-NN technique operates by defining an appropriate metric in the color and
magnitude multidimensional space, and determines a ``score'' for each object,
based on its ``distance'' from the colors and magnitudes of known T dwarfs.
This method has been successful in identifying brown dwarf candidates in other
$Spitzer$ IRAC data, which were then confirmed with proper motion analysis
(Luhman et al. 2007).  However, it is unreliable at identifying substellar
candidates when spectral type is earlier than T3.

We required a $\geq$ 5$\sigma$ detection in all four IRAC channels in order for a source to be considered a viable candidate. No four-channel candidates  exhibited colors and magnitudes consistent with a brown dwarf spectral type of T3 or later.    The four-channel $k$-NN analysis was originally performed on all targets, before the second-epoch observations were carried out, and the more powerful common proper motion analysis became possible.

\section{Observational Sensitivities}
Our sensitivity to companions at narrow separations is limited
most strongly by the primary star flux.  At wider separations, the
sensitivities are limited by a combination of incident field stars, zodiacal
flux levels, and photon noise.    For separations from the primary of
30$\arcsec$ and greater, we estimated sensitivities based on the standard
deviations of flux levels from the frames in the dither pattern.  Since we are
most interested in a statistical representation of sensitivity levels, we did
not measure wide-separation sensitivities for all 117 images, but rather determined flux level deviations for a representative sample (covering a range of zodiacal levels) and then extrapolated to the complete set.  

Because we have difficulty detecting companions at the positions of field
stars, we assumed that no brown dwarf identification could be made within two
FWHM of a field star.  We used this approach to determine a reasonably
accurate incompleteness fraction that could be incorporated into the
subsequent Monte Carlo simulations.  At narrow separations ($<$30$\arcsec$)
from the primary, we used only the final combined image, and estimated
sensitivities based on the standard deviation of flux inside concentric annuli
centered on the primary star.  The thickness of each annulus was one FWHM.  It
is likely that this method of determining high-contrast sensitivity
over-estimates the noise somewhat, since asymmetries in the primary star PSF
may be interpreted as noise.  For the purposes of our statistical analysis
however, these sensitivity limits served as a quantifiable and useful tool to give us a conservative measurement of our ability to identify low mass companions at narrow separations.

Figure 2 shows our sensitivity curves, as compiled from the survey's median,
best 15\%, and worst 15\% of observations.  Best and worst levels are  based
on measured sensitivities between 8 and 15 arcseconds from the primary and are
affected by primary star brightness, field star density, and zodiacal
background levels.  In the cases of two-epoch data when one epoch had better
sensitivities than the other, we adopted the sensitivity levels of the worse epoch.  For the median curve, the 5$\sigma$ limiting magnitudes imply a sensitivity to substellar companions $\sim$600K or warmer, for separations $>$7$\arcsec$ (assuming brown dwarf temperature-magnitude relations from Golimowski et al. 2004 and Patten et al. 2006). 

\section{Monte Carlo Population Simulations}
To derive companion statistics from the observational data, we used Monte
Carlo simulations combined with Bayesian modeling as described in Carson et
al. (2006).  For 2MASS J07171626+5705430, DENIS	J153941.8-052042, and 2MASS
J17281150+3948593,  where brown dwarf companion identifications were carried
out using a color-color analysis, instead of common proper motions, we used
the 5.8 $\mu$m final images, which were the least sensitive of the four and
therefore dictated the achievable sensitivities for that analysis.  For all
other targets, we generated sensitivity maps from the 4.5 $\mu$m reduced
images using the protocol described in the previous section.  The sensitivity
maps served as inputs to the Monte Carlo simulations.  Following Carson et
al. (2006), we assumed, as a baseline configuration, that potential companions
have a random eccentricity (between 0 and 0.9), random inclination, and random
longitude of pericenter.  The Monte Carlo simulations use discrete steps to
cover the range of possible companion orbital parameters.  In our case,  we
sampled semimajor axes of 10, 20, 30, 40, 80, 100, 200, 400, 600, 800, 1000,
2000, 4000, 6000, and 10000 AU, and companion temperatures of 525, 633, 850, 950, and 1100 K.  We chose these values because they provided a balance between a wide range of parameter space and an adequately small step size.   

\subsection{Detection Probabilities}

Figure 3, generated from the Monte Carlo simulations, displays the likelihood
of our detecting a substellar companion of a given temperature around a
typical target (if a companion does indeed exist) as a function of semimajor
axis.  These curves take into account orbital projection effects (including
the variations in time spent at different parts of the orbit) as well as
survey sensitivities.  For this plot, we  assume a conversion between brown
dwarf temperature and IRAC magnitude using relations in Patten et al. (2006)
and Golimowski et al. (2004); Patten et al. (2006) allows for the conversion
from IRAC magnitude to spectral subtype and Golimowski et al. (2004) allows
for the conversion from spectral subtype to temperature.  Figure 3 indicates
that a 600-1100K companion with a semimajor axis somewhere between 35 and 1200
AU has a 60\% chance of being detected in a given target observation, while a
500-600K companion between 60 and 1000 AU has a 16\% chance of being detected.
For the displayed plot, we selected the two temperature bins in order to present
effective sensitivities for both relatively bright brown dwarfs, as well as fainter brown
dwarfs that lie near the fringes of our sensitivity limits.  The semimajor
axis boundaries, quoted above, correspond to points on the curve where detection probabilities drop
to around 50 to 60\% of the peak probability levels.    

Figure 4 shows the detection probabilities for a combined 500-1100K sample
companion population (uniformly distributed across temperature).  The plot
also displays how detection probabilities are affected if the hypothetical
population is skewed toward circular orbits or highly eccentric orbits.  A
highly eccentric population translates into a sensitivity to smaller semimajor
axes than for the case of more circular orbits.  This is due to the fact that
an orbiting companion with a high eccentricity spends a proportionately larger
fraction of its orbit at separations exceeding the semimajor axis.  

For the brown dwarf temperature and semimajor axis range that we are
considering, there is no observational evidence, either in our data or in the
published literature, that suggests a variation in the relative frequency of companions as a
function of temperature.  Because of this unknown, we assume a flat
distribution of brown dwarf companion frequency versus temperature.  However,
Figure 5 displays detection probabilities for a combined 500-1100K sample, if
we instead assume that the relative frequency of brown dwarf companions with
temperature mimics that of the field brown dwarf distribution, as concluded by
Bayesian analyses in Allen et al. (2005).  In this modified case, the peak detection probability decreases by about 10\%, owing to
the input of a brown dwarf distribution where colder
companions are more frequent than warmer ones.

\subsection{Population Upper Limits}   

Following the Bayesian analysis described in Carson et al. (2006), a companion fraction upper limit was derived from the detection probability curves.  Figure 6 shows 90\% certainty upper limits to the fraction of systems with 600-1100K and 500-600K companions.  The plot indicates that, at a 90\% confidence level, at most 3.4\% of survey targets have an undetected 600-1100K companion between 35 and 1200 AU.  At most 12.4\% of survey targets have an undetected 500-600K companion between 60 and 1000 AU (these upper limit values translate to 4.4\% and 16.1\%, respectively, for a 95\% certainty calculation).  

We decided to report the companion statistics for a constrained
temperature range, as opposed to a mass range, because: 1) temperature
is more directly observable than mass, and 2) the target sample contains
 biases which more strongly affect mass-constrained statistics than
temperature-constrained statistics.  Since brown dwarfs of a given mass grow
fainter with age, Monte Carlo simulations that consider companion mass must incorporate theoretical evolutionary models as well as a-priori information on
target system age.  Most of the target stars in our sample have relatively
unknown ages.   

As described in Section 2, the target list contains known biases against
small-separation ($\leq$6$\arcsec$) stellar binaries.  In the case of
temperature-constrained statistics, all such deselected systems have companion
temperatures greater than the 500-1100K temperature
range, and therefore do not affect
derived statistics.  But it is possible for a
$\leq$6$\arcsec$ companion to affect mass-constrained statistics,
since an unknown target system age implies a possible substellar mass
overlapping with those explored by the Monte Carlo simulations.  See Baraffe
et al. (2003) for a discussion of the relationship between brown dwarf mass,
age, and temperature.

The published substellar companions described in Table 1 come from a
range of surveys and detection techniques (from Doppler spectroscopy
to adaptive optics imaging) and hence contain a variety of detection biases.
All of the published companions either have semimajor axes outside of those
explored by the Monte Carlo simulations, or have temperatures outside of the
500-1100K range examined by our survey.  Therefore, they do not affect
temperature-constrained companion statistics.  But due to the unknown ages of
survey targets, some published companions, with semimajor axes probed by our
Monte Carlo simulations, may have masses that are relevant to mass-constrained, companion statistics.  Proper inclusion of
such cases in a statistical analysis would therefore require a consideration
of all published detections' observational biases.  Such considerations are beyond the
scope of this paper.

\subsection{Comparison With Other Surveys}
A wide variety of observational surveys have probed the question of multiplicity among low-mass stars and brown dwarfs (Burgasser et al. 2007).  A recent statistical analysis on companions to low-mass stars and brown dwarfs is the investigation of Allen (2007).  Taking into account observational biases and projection effects, that analysis combines the observational data from many different programs to conclude that the wide ($\gtrsim$20 AU) companion fraction to dwarfs of spectral-type M6 and later is no more than 1\%-2\%.  This assumes a completeness to mass ratios as small as 0.5-0.7.  As mentioned earlier, our analysis is defined by companion temperature, rather than mass.  Hence, a direct comparison is difficult.  However, around a typical target in our sample, the sensitivity floor of $\sim$500K translates to an expected minimum mass ratio $\sim$0.3 (assuming Baraffe et al. 2003 evolutionary models and Allen et al. 2005 brown dwarf spectral-type/mass/age statistics).  Hence, our companion fraction upper limit is generally consistent with the Allen (2007) result, and likely extends the wide-separation companion paucity to noticeably smaller companion masses and brightnesses.

\section{Summary}
Using IRAC on the \textit{Spitzer Space Telescope}, we have conducted a
substellar companion imaging survey of 117 nearby (median distance $\sim$ 10
pc) M, L, and T dwarf systems.   Using 4.5 $\mu$m imaging observations with
multi-epoch common proper motion tests for the majority of targets, we
achieved  sensitivities to substellar companions as cool as $\sim$500K  for
semimajor axes between 60 and 1000 AU;  for $>$600K  companions, we achieved sensitivities to  semimajor axes ranging from
35 to 1200 AU.  Our survey discovered no new companions.  Based on the
observational results, Monte Carlo population simulations determined that for
the given target sample, the 600-1100K companion fraction is $<$ 3.4\% for
semimajor axes of 35-1200 AU; the 500-600K companion fraction is $<$ 12.4\%
for semimajor axes of 60-1000 AU.  We note that these Monte Carlo results
represent the first rigorous statistics, among any parent star target class, of the wide separation ($\geq$60 AU) $<$600K companion population.  

The statistical results are consistent with previous studies that report a scarcity of wide-separation ($>$30 AU) substellar companions to low mass stars and brown dwarfs (e.g. Allen 2007, Burgasser et al. 2007).  Our Monte Carlo results complement those previous studies by expanding companion statistics to include considerably lower temperature companions, for the given orbital separations.  The Monte Carlo results are also consistent with brown dwarf formation models, such as ejection scenarios (see Section 1), that      
predict a scarcity of the widest and least massive binary systems.  While our results are not comprehensive enough to definitely prove such theories, the unprecedented sensitivity and sample size will provide valuable constraints for the current and next generation of brown dwarf formation theories.  Our study represents a constrained subset of a larger $Spitzer$ IRAC companion search effort, which will be reported in a future publication.






\acknowledgments
We thank our referee, Angelle Tanner, for helpful comments which improved the
manuscript.  This work is based on observations made with the \textit{Spitzer Space Telescope}, which is operated by the Jet Propulsion Laboratory (JPL), California Institute of Technology under a contract with NASA.  Support for part of this work was provided by NASA through an award issued by JPL.  Support for the IRAC instrument was provided by NASA under contract number 1256790 issued by JPL.  J. C. C. was supported by grant AST-1009203 from the National Science Foundation.  K. L. was supported by grant AST-0544588 from the National Science
Foundation. The Center for Exoplanets and Habitable Worlds is supported
by the Pennsylvania State University, the Eberly College of Science, and the
Pennsylvania Space Grant Consortium.  This research has made use of the following databases: 1) the M, L, and T dwarf compendium housed at DwarfArchives.org and maintained by Chris Gelino, Davy Kirkpatrick, and Adam Burgasser; 2) the SIMBAD database, operated at CDS, Strasbourg, France; 3) the NASA/IPAC/NExScI Star and Exoplanet Database, which is operated by JPL; and 4) the Washington Double Star Catalog maintained at the U.S. Naval Observatory.



{\it Facilities:} \facility{Spitzer (IRAC)}.

\clearpage



\begin{figure}
\epsscale{1.0}
\plotone{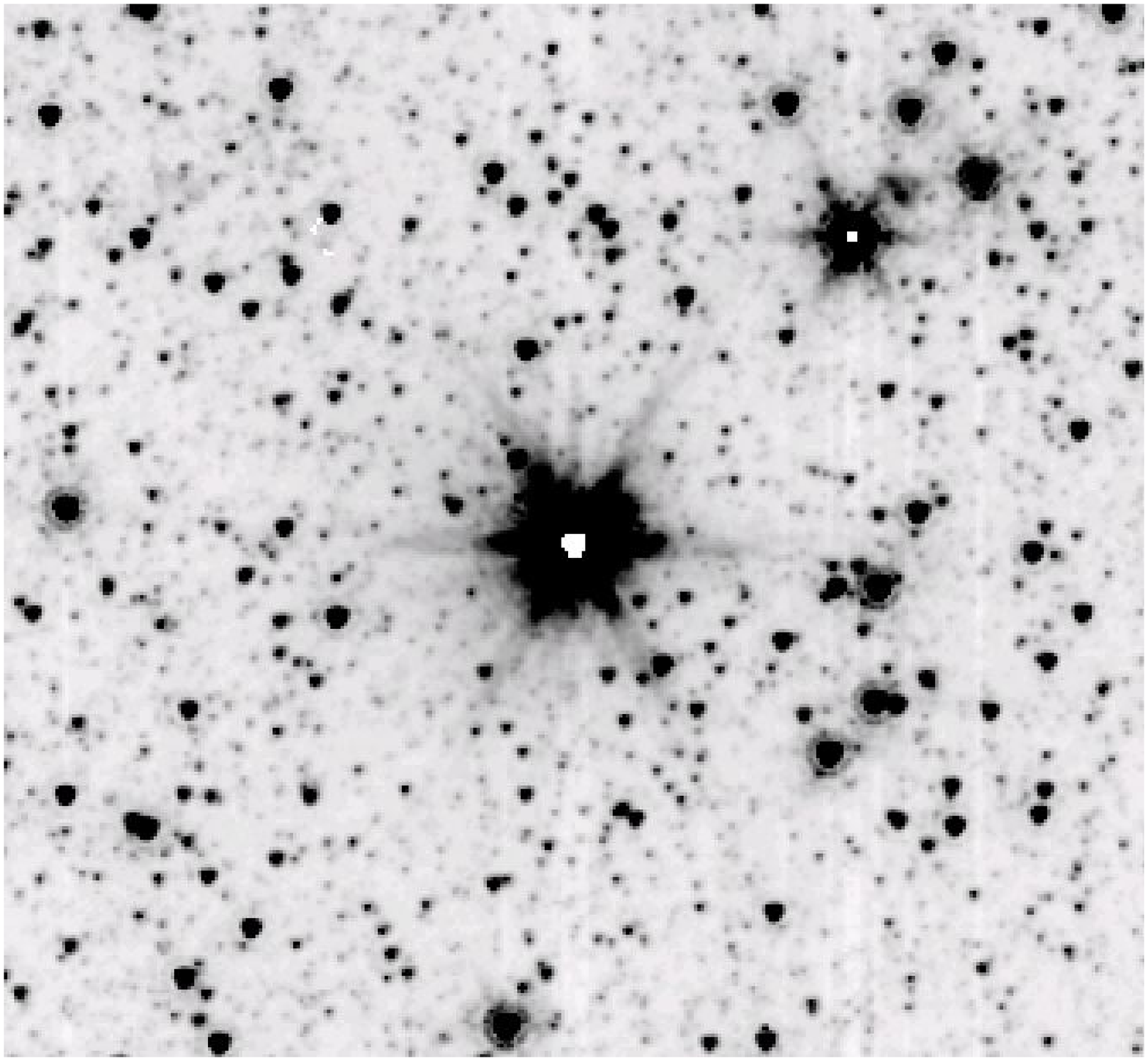}
\caption{Reduced \textit{Spitzer} IRAC 4.5 $\mu$m image of GJ 551 (one of the
  brighter targets).  Displayed field of view is $\sim$ 4.$\arcmin$8 $\times$
  4.$\arcmin$8.  The white spot at the center of the bright star is due to saturation.  The image illustrates how sensitivities to companions are affected by primary star brightness and incident field stars.}
\end{figure}

\clearpage

\begin{figure}
\epsscale{1.0}
\plotone{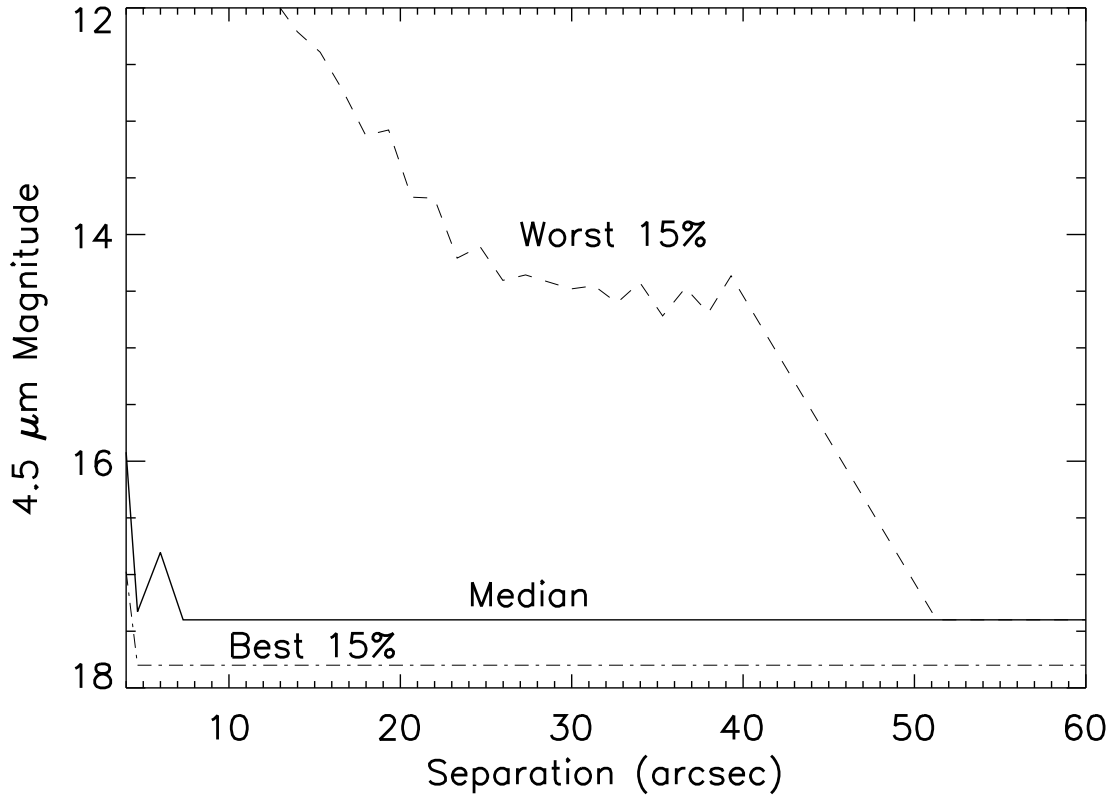}
\caption{The 4.5 $\mu$m sensitivity curves displaying limiting magnitude as a
  function of separation from the primary.  The solid curve represents median
  survey sensitivities.  The dashed curve represents median sensitivities for
  the worst 15\% of observations.  The dot-dashed curve represent median
  sensitivities for the best 15\% of observations.  All minimum magnitudes
  correspond to 5$\sigma$ detections.  Outside of the halo of the primary,
  typically at $\sim$ 7$\arcsec$ separations, sensitivity estimates are purely a function of zodiacal background levels and field star density.}
\end{figure}

\clearpage

\begin{figure}
\epsscale{1.0}
\plotone{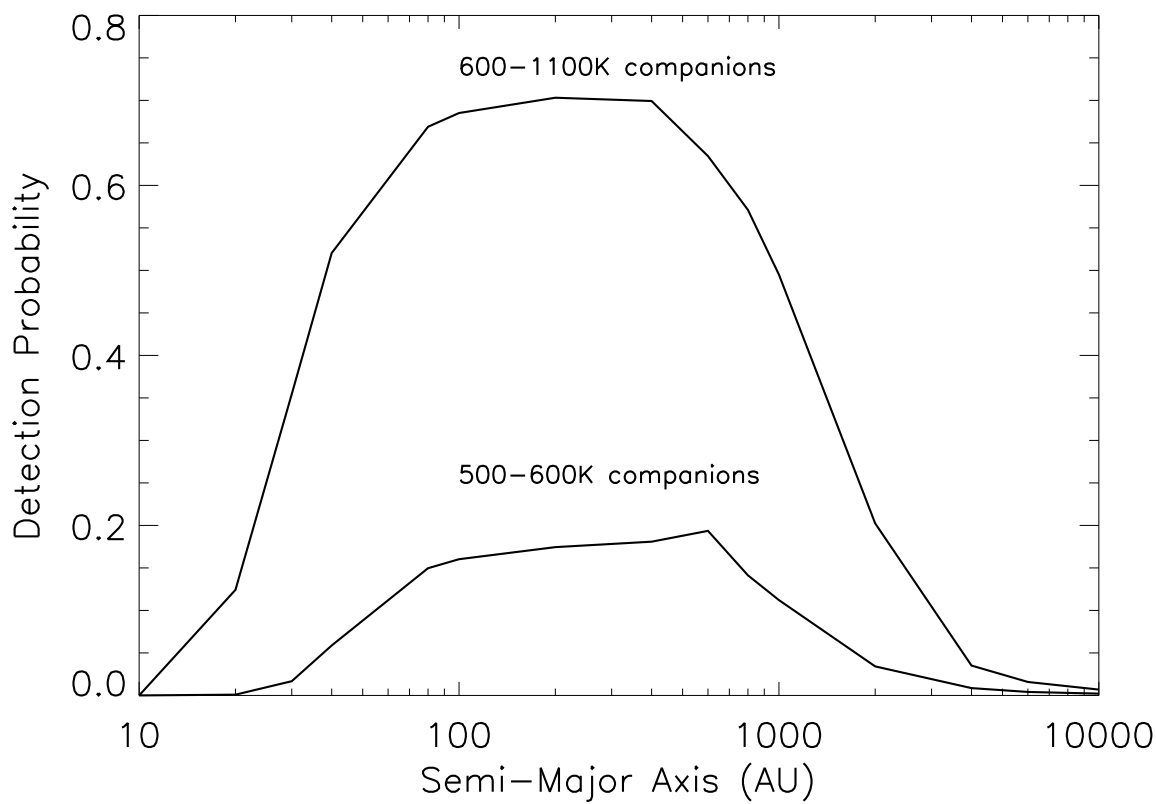}
\caption{Detection probability versus semimajor axis for substellar
  companions, as output from Monte Carlo simulations.  The plotted detection
  probability represents the likelihood that the survey successfully
  identifies a substellar companion, if one does indeed exist, as a function
  of companion semi-major axis.  These curves represent an average over all survey observations.  
The top curve is the detection probability for 600-1100K companions and the bottom curve is the probability for 500-600K companions.}
\end{figure}

\begin{figure}
\epsscale{1.0}
\plotone{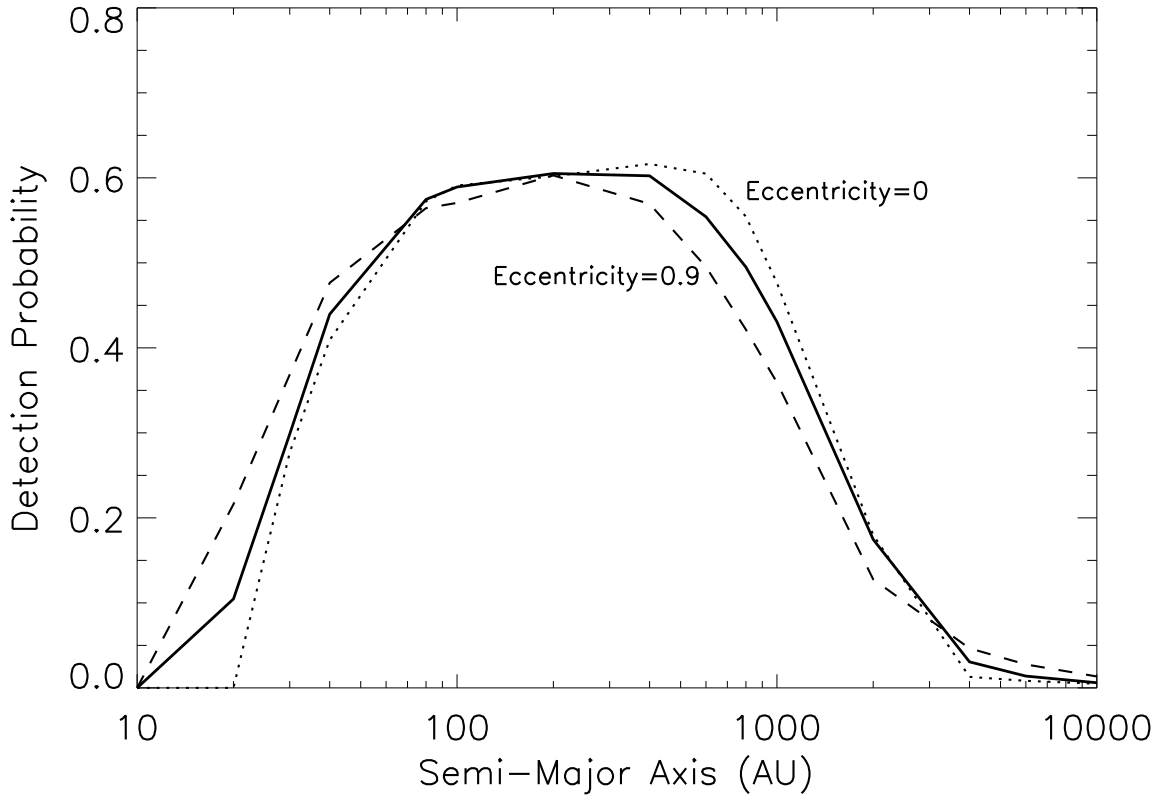}
\caption{Detection probability  versus semimajor axis for substellar companions, as output from Monte Carlo simulations.  The solid curve indicates the detection probability for a uniform 500-1100K sample companion population with random eccentricity. (It is effectively a consolidation of the Figure 3 curves.)  The dashed and dotted curves demonstrate how detection probabilities change when one assumes a highly eccentric ($e$=0.9) or highly circular ($e$=0) orbital population.  Note that a highly eccentric companion population implies a sensitivity to narrower semimajor axes than a more circular companion population.  }
\end{figure}

\begin{figure}
\epsscale{1.0}
\plotone{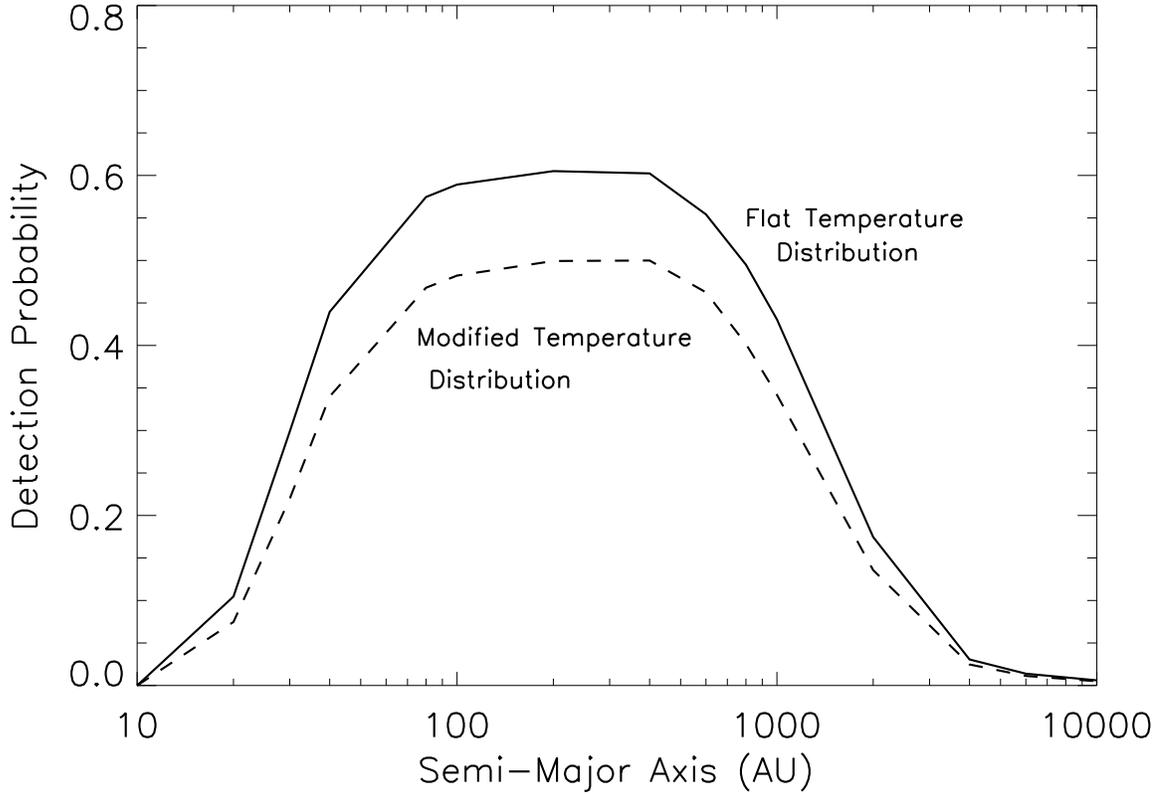}
\caption{Detection probability  versus semimajor axis for substellar companions, as output from Monte Carlo simulations.  The solid curve indicates the detection probability for a uniform 500-1100K sample companion population where the relative frequency of brown dwarf companions varies flatly with temperature. (It is effectively a consolidation of the Figure 3 curves.)  The dashed curve demonstrates how detection probabilities change when one assumes that the relative frequency of companion brown dwarfs versus temperature mimics closely that of field brown dwarf populations.  Detection probabilities drop about 10\% for the modified temperature distribution, owing to the higher relative frequency of colder, and therefore fainter, companions.  }
\end{figure}

\begin{figure}
\epsscale{1.0}
\plotone{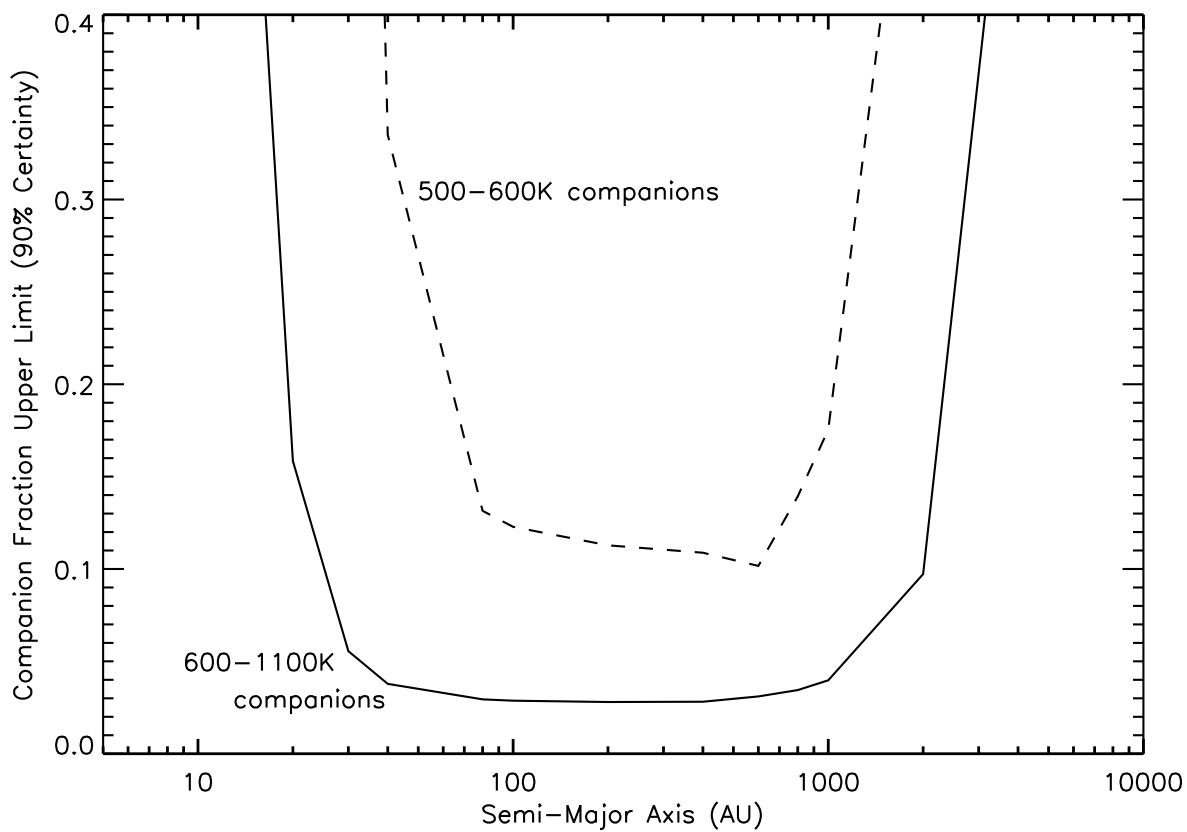}
\caption{Upper limit for the frequency of
brown dwarf companions, at a 90\% confidence level, versus semimajor axis, as outputted by our Monte Carlo simulations.  The upper curve shows statistical upper limits for 500-600K companions and the bottom curve shows upper limits for 600-1100K companions.  The curves indicate that at most 3.4\% of survey targets have an undetected 600-1100K companion between 35 and 1200 AU.  At most 12.4\% of survey targets have an undetected 500-600K companion between 60 and 1000 AU.      
 }
\end{figure}

\clearpage

\begin{deluxetable}{ccccp{5cm}cp{5cm}cccc}

\tabletypesize{\scriptsize}
\rotate
\tablecaption{Sample of Late-M, L, and T Dwarf Systems\label{tbl-1}}
 \tablehead{
\colhead{Name} & $Spitzer$ PID & \colhead{Spectral} & \colhead{Distance\tablenotemark{a}} &
\colhead{Multiplicity} & \colhead{Spectral Type Ref.} & \colhead{} &
\colhead{} & \colhead{} &
\colhead{} & \colhead{} \\
\colhead{}  & Number & \colhead{Type} & \colhead{(pc)} &
\colhead{Notes\tablenotemark{b}} & \colhead{[\& Multiplicity Ref.]} & \colhead{} &
\colhead{} & \colhead{} &
\colhead{} & \colhead{} \\
}
\startdata
GJ 825       &33, 30298&    M0  &    4.0  &      & 1& & & & & \\
GJ 412 AB       &33, 30298 &     M0.5+M5.5  &       4.8   &    binary; separation = 28$\arcsec$ (resolved) & 2 [3,4] &  & & & & \\
GJ 229 AB      & 35, 30298&   M1+T7 &  5.8   &    binary;  separation = 7.$\arcsec$8 & 5 [5] & & & & & \\
GJ 191       &33, 30298&    M1.5 &     3.9      &  &1 & & & & & \\
GJ 15 AB     & 33, 30298   &     M1.5+M3.5  &        3.6    &   binary; separation = 34.$\arcsec$8 (resolved) & 1 [1,4]&  & & & & \\
GJ 887     &   33, 30298  & M1.5  &    3.3    &    & 1 & & & & & \\
GJ 411      &33, 30298&     M2  &    2.6      &  &  1& & & & & \\
GJ 752 AB        &35, 30298&     M2+M8   &  5.9   & binary;   separation = 74" (resolved) & 2 [3,4] &  & & & & \\
GJ 1         &33, 30298&   M3   &   4.4      &  & 1 & & & & & \\
GJ 388       &33, 30298&    M3  &   4.7      &  &  1& & & & & \\
GJ 628   &  33, 30298  &    M3   &   4.3 &  & 1 & & & & & \\
GJ 674      &33, 30298  &    M3    &  4.5   &    known exoplanet from Doppler spectroscopy & 1 [1] & & & & & \\
GJ 687    &33, 30298&      M3   &   4.5   &    & 1 & & & & & \\
GJ 832       & 33, 30298 & M3  &    4.9  &     known exoplanet from Doppler spectroscopy & 1 [1] & & & & & \\ 
GJ 860 AB       &33, 30298&     M3+M4    &     4.0     &  binary; separation = 3.$\arcsec$4 & 1 [1,6] & & & & & \\
GJ 725 AB  &  33, 30298   &     M3+M3.5  &       3.6  &     binary; separation = 13.$\arcsec$3 (resolved) & 1 [1] & & & & & \\
GJ 729      &33, 30298&     M3.5   &   3.0   &     & 1& & & & & \\
GJ 273      & 33, 30298&    M3.5    &  3.8      &  & 1& & & & & \\
GJ 876        &  33, 30298  & M3.5   &   4.7  &     3 known exoplanets from Doppler spectroscopy&1 [1] &  & & & & \\
GJ 1001 A    & 35, 30298     &M3.5 & 9.5\tablenotemark{c} & part of a triple system;  18.$\arcsec$2 (resolved) from BC & 2 [7] &  & & & & \\
GJ 447      &33, 30298 &     M4   &   3.3      &  &  1& & & & & \\
GJ 699      &33, 30298&     M4  &    1.8      &  & 1 & & & & & \\
GJ 54.1       &33, 30298&   M4.5  &    3.7      &  &1 & & & & & \\

GJ 83.1    &33, 30298&    M4.5   &   4.5      &  & 1 & & & & & \\
GJ 234 AB   &  33, 30298 &      M4.5+??    &        4.1  &     binary; separation = 1.$\arcsec$3  &1 [1,4]  & & & & & \\
GJ 866 ABC       & 33, 30298 &    M4.5+M5+M6\tablenotemark{d}    &    3.3 &  triple system; AC to B separation = 0.$\arcsec$35; A to C separation $<$ 0.$\arcsec$1 & 8,9 [8] & & & & & \\
GJ 1093     & 35, 30298  &  M5  &      7.8    &     & 2 & & & & & \\
GJ 1156  &35, 30298&   M5.5  &    6.5   &      & 2 & & & & & \\
GJ 1002      &33, 30298&     M5.5   &   4.7    &    & 2 & & & & & \\
LHS 288        &   33, 30298  &   M5.5  &    4.5  &      & 2 & & & & & \\
GJ 473 AB  &      33, 30298    & M5.5+??      &   4.4  & binary; separation = 0.$\arcsec$7  & 1 [1,4]& & & & & \\
GJ 551      & 33, 30298&    M5.5     &    1.3   &  part of a triple system; 16.$\arcmin$7 separation from a binary (outside field of view) & 1 [10] & & & & & \\
GJ 905     &  33, 30298   & M5.5   &   3.2     &   & 1& & & & & \\
GJ 1061    &  33, 30298    &M5.5   &   3.7      &  & 1 & & & & & \\
GJ 1245 ABC      & 33, 30298&     M5.5+M5.5+$>$M6  &    4.7 &  triple system; AB separation = 7.$\arcsec$0 (resolved); AC separation = 0.$\arcsec$6. & 11 [11] & & & & & \\
GJ 65 AB     &33, 30298 &     M5.5+M6  &        2.7    &   binary; separation = 0.$\arcsec$7 & 1 [1,4]  && & & & \\
GJ 406     &33, 30298&     M6  &    2.4      &  &  1& & & & & \\
GJ 1111       &33, 30298 &  M6.5     & 3.6   &     & 2 & & & & & \\
LHS 292          &33, 30298&    M6.5  &    4.6   &     & 2 & & & & & \\
SO 025300.5+165258    & 33, 50013 &     M7  &      3.7   &     & 2 & & & & & \\
LHS 3003        &35, 30298&    M7    &    6.4    &    & 2 & & & & & \\
GJ 644 C       &35, 30298&     M7    &    6.5   &  part of a triple system; C to A separation = 4.$\arcmin$9 (outside field of view)  & 2 [3,4] &  & & & & \\     
LHS 132       &35, 30298&   M8   &    11.0\tablenotemark{c}    &   & 2 & & & & & \\
LHS 2021         &35, 30298&     M8  &      15.0\tablenotemark{c}   &    & 2 & & & & & \\
2MASS J18353790+3259545  &  35, 30298   &     M8.5  &    5.7    &    & 2 & & & & & \\
2MASS J03202839-0446358      & 35, 30298 &    M8.5+T5    &   26.0\tablenotemark{c}   & binary;  separation $<$ 0.$\arcsec$33  & 12 [12] & & & & & \\
LP 944-020        & 33, 30298 &   M9     &   5.0    &    & 2 & & & & & \\
LHS 2065           &  35, 30298&  M9   &     8.6    &   & 2 & & & & & \\
LHS 2924           &  35, 30298  &M9    &    11.0   &    & 2 & & & & & \\
DENIS	J104814.6-395606     &33, 30298 &      M9    &  4.5\tablenotemark{e}      &  & 3 & & & & & \\
DENIS	J002105.8-424442      & 35, 30298 &     M9.5  &    24.9\tablenotemark{c}   &    & 2 & & & & & \\
BRI 0021-0214      &  35, 30298   &M9.5    &  11.5   &    & 2 & & & & & \\
2MASS J12043036+3212595     &   35, 70021  & L0    &    43.8\tablenotemark{c}    &   & 2 & & & & & \\
2MASS J04510093-3402150      & 35, 60046&    L0.5    &  28.0\tablenotemark{c}     & & 2 & & & & \\
2MASS J07464256+2000321 AB  & 35, 30298     &     L0.5+L  & 12.2   &   binary; separation = 0.$\arcsec$22  & 13 [13] & & & & & \\
Kelu-1     & 35, 30298 &   L2+L3.5  & 18.5  & binary; separation = 0.$\arcsec$37    & 14 [14] & & & & & \\
2MASSW J1300426+191235 	      & 35, 30298 &    L1     &   14.6\tablenotemark{c}    &   & 2 & & & & & \\
2MASS J14392837+1929150      &  35, 30298  &  L1     &   14.3  &     & 2 & & & & & \\
2MASS J15551573-0956055       & 35, 30298&    L1    &    13.8\tablenotemark{c}    &   &  2& & & & & \\
2MASS J16452211-1319516       &  35, 30298  &L1.5    &  13.3\tablenotemark{c}   &    & 2& & & & & \\
2MASS J10170754+1308398      &  35, 60046&   L2+earlyL    &    30.9\tablenotemark{c}   & binary;  separation = 0.$\arcsec$10   & 15 [15]& & & & & \\
2MASS J11553952-3727350     &35, 30298 &     L2     &   17.1\tablenotemark{c}     & & 2& & & & \\
DENIS	J105855.9-154814 	     &   35, 50013 &    L3     &   17.2   &    &2 & & & & & \\
2MASS J1506544+132106      & 35, 30298&     L3   &     11.4\tablenotemark{c}   &    & 2& & & & & \\
2MASS J1721039+334415      & 35, 30298&     L3   &     15.9\tablenotemark{c}    &   &2 & & & & & \\
SDSS J202820.32+005226.5     &35, 60046&     L3    &    19.0\tablenotemark{c}    &   & 2& & & & & \\
2MASS J21041491-1037369     & 35, 30298&    L3    &    15.7\tablenotemark{c}   &    &2 & & & & & \\
2MASS J00361617+1821104    & 35, 30298   &   L3.5  &    8.8  &      & 2& & & & & \\
DENIS	J153941.8-052042        & 35, 30298 &   L4     &   16.0\tablenotemark{c}   &    &2 & & & & & \\
2MASS J01410321+1804502        &  35, 30298&   L4.5  &    16.7\tablenotemark{c}   &    & 2& & & & & \\
2MASS J06523073+4710348       &35, 60046 &    L4.5  &    8.8\tablenotemark{c}     &   & 2& & & & & \\
2MASS J22244381-0158521       & 35, 30298 &   L4.5  &    11.4    &   & 2& & & & & \\
SDSS J053951.99-005902.0       & 35, 30298     &L5   &     13.2    &   &2 & & & & & \\
2MASS J08354256-0819237     & 35, 30298 &    L5   &     6.7\tablenotemark{c}   &     &2 & & & & & \\
2MASS J09083803+5032088    & 35, 30298  &    L5    &    14.0\tablenotemark{c}    &   &2 & & & & & \\
2MASS J15074769-1627386      & 35, 3736&    L5   &     7.4     &   & 2& & & & & \\
GJ 1001  BC    & 35, 30298    & L5+midL   &     9.5   & binary;  part of a triple system; B to C   separation = 0.$\arcsec$09; BC to A separation = 18.$\arcsec$2 (resolved)  & 7 [7] &  & & & & \\ 
SDSS J133148.92-011651.4         & 35, 30298&   L6    &    26.3\tablenotemark{c}   &    &2 & & & & & \\
2MASSW J1515008+484742      &35, 30298 &     L6    &    12.3\tablenotemark{c}   &    & 2& & & & & \\
2MASS J07171626+5705430         & 35, 30298& L6.5   &   16.8\tablenotemark{c}    &   & 2& & & & & \\
SDSS J042348.57-041403.5    &   35, 30298  &   L6.5+T2   &  15.2 &     binary; separation = 0.$\arcsec$16 &16 [16] & & & & & \\
2MASS J15261405+2043414      &35, 30298&     L7     &   25.2\tablenotemark{c}    &   & 2& & & & & \\
2MASS J17281150+3948593       & 35&    L7+L/T  &   24.4   &  binary; separation = 0.$\arcsec$13 & 17 [17] & & & & & \\
2MASS J08251968+2115521    &35, 30298&      L7.5  &    10.6   &    &2 & & & & & \\
DENIS	J025503.3-470049     &35, 30298 &    L8   &     7.2\tablenotemark{c}      &  &2 & & & & & \\
SDSS J085758.45+570851.4     & 35, 30298 &      L8  &      13.3\tablenotemark{c}     &  & 2& & & & & \\
GJ 337 CD       & 35, 30298 &    L8+$\sim$L8   &  20.4  &    part of a quadruple  system; CD  separation = 0.$\arcsec$53;    AB to CD separation     = 43$\arcsec$ (resolved)    & 18 [18]&  & & & & \\
2MASS J16322911+1904407   & 35, 30298 &      L8   &     15.2     &  &2 & & & & & \\
2MASS J05325346+8246465   &   35, 30298  &    sdL    &   8.4\tablenotemark{c}      &  & 2 & & & & & \\
SDSS J015141.69+124429.6    &35, 30298&      T0.5    &  21.3     &  &2 & & & & & \\
SDSS J083717.22-000018.3     &35, 50013&     T1   &     29.4     &  &2 & & & & & \\
$\epsilon$ Ind BC  &  33, 60046   &     T1+T6   &  3.6  & binary; part of a   triple system; BC   separation = 0.$\arcsec$73;   BC to A    separation = 6.$\arcmin$7 (outside field of view)  & 2 [19,20] & & & & & \\
SDSS J125453.90-012247.4    &35, 3736&      T2   &     11.8     &  & 2& & & & & \\
SDSS	J102109.69-030420.1     &35, 50013&   $\lesssim$T2+T5  &   29.4 & binary;    separation = 0.$\arcsec$17 & 16 [16]& & & & & \\
SDSS J175032.96+175903.9    &35, 50013&      T3.5  &    27.8     &  &2 & & & & & \\
2MASS J22541892+3123498 	     &35, 30298&      T4     &   15.0\tablenotemark{c}     &  &2 & & & & & \\
SDSS	J020742.87+000056.0     &35, 50013&      T4.5    &  28.6     &  & 2& & & & & \\
2MASS J05591914-1404488      &35, 30298&     T4.5    &  10.2     &  &2 & & & & & \\
SDSS J092615.38+584720.9        & 35, 30298&   T4.5\tablenotemark{f}  & 17.0\tablenotemark{c}  &    binary;   separation = 0.$\arcsec$07& 16 [16] & & & & & \\
2MASS J07554795+2212169    & 35, 30298&      T5   &     13.6\tablenotemark{c}     &  & 2& & & & & \\
2MASS J23391025+1352284      &35, 30298 &     T5   &     15.5\tablenotemark{c}     &  &2 & & & & & \\
2MASSW J2356547-155310     &35, 30298 &     T5    &    14.5     &  &2 & & & & & \\
2MASS J15344984-2952274       &35, 30179&     T5.5\tablenotemark{f}  & 13.5   &   binary;    separation = 0.$\arcsec$07& 21 [21]& & & & & \\
2MASS J15462718-3325111     & 35, 60046&      T5.5   &   11.4     &  & 2& & & & & \\
SDSS J111010.01+011613.1    & 35, 30298&       T5.5   &   14.2\tablenotemark{c}     &  &2 & & & & & \\
2MASS J02431371-2453298   & 35, 30298&      T6      &  10.6     &  &2 & & & & & \\
2MASS J12255432-2739466   & 35, 30298 &      T6+T8  &   13.3     & binary;  separation = 0.$\arcsec$28 &  21 [21]&  & & & & \\
SDSS J162414.37+002915.6   &  35, 30298&       T6      &  11.0     &  & 2& & & & & \\
2MASS J09373487+2931409 	      &35, 30298   &  T6p   &    6.1      &  & 2& & & & & \\
2MASSW J1047539+212423 	        & 35, 30298 & T6.5  &    10.5     &  & 2& & & & & \\
2MASS J12373919+6526148    &  35, 30298 &      T6.5  &    10.4     &  &2 & & & & & \\
SDSS J134646.45-003150.4    & 35, 30298  &    T6.5  &    14.7     &  & 2& & & & & \\
2MASS J15530228+1532369   &  35, 30298  &     T6.5+T7  &   9.2\tablenotemark{c} &      binary;     separation = 0.$\arcsec$35 & 16 [16]& & & & & \\
2MASS J07271824+1710012    & 35, 30179  &     T7   &     9.1      &  &2 & & & & & \\
2MASS J12171110-0311131    &  35, 30298 &     T7.5  &    11.0     &  &2  & & & & & \\
GJ 570 D     &  35, 30179 &   T7.5   &   5.9      &  & 2& & & & & \\
2MASS J04151954-0935066 &  35, 30179  &      T8    &    5.8      &  & 2& & & & & \\


\enddata
\tablecomments{Targets are sorted according to spectral type.  The IRAC magnitudes of all the L and T-type targets, and most of
  the M-type
  targets, are published in Patten et al. (2006).}
\tablenotetext{a}{Values from Patten et al. (2006), when available, or NASA/IPAC/NExScI Star and Exoplanet Database (\textit{NStED}; http://nsted.ipac.caltech.edu), unless otherwise noted.}
\tablenotetext{b}{All multiple systems are unresolved unless otherwise noted.}
\tablenotetext{c}{Distance approximated from 3.6 $\mu$m magnitude, spectral type, and magnitude-spectral-type relations from Patten et al. (2006).}
\tablenotetext{d}{Spectral types derived by authors using Delfosse et al. (1999) mass estimates and Baraffe \& Chabrier (1996) mass/spectral-type relations.}
\tablenotetext{e}{From Neuh{\"a}user et al. (2002).}
\tablenotetext{f}{Composite identification.}
\tablerefs{
(1) NASA/IPAC/NExScI Star and Exoplanet Database,
  http://nsted.ipac.caltech.edu; (2) Patten et al. 2006; (3) Montes et
  al. 2001; (4) Washington Double Star Catalog; Mason et al. 2001;  (5)
  Nakajima et al. 1995; (6) Hipparcos; Perryman et al. 1997; (7) Golimowski et
  al. 2004; (8) Delfosse et al. 1999;  (9) Baraffe \& Chabrier 1996;  (10)
  Wertheimer \& Laughlin 2006; (11) Schroeder et al. 2000; (12) Burgasser et
  al. 2008; (13) Reid et al. 2001; (14) Gelino et al. 2006; (15) Bouy et al. 2003; (16) Burgasser et al. 2006; (17) Gizis et al. 2003; (18) Burgasser, Kirkpatrick, \& Lowrance 2005; (19) McCaughrean et al. 2004; (20) Scholz et al. 2003;  (21) Burgasser et al. 2003.   }
\end{deluxetable}





\end{document}